\documentclass[vecphys]{svmult}

\usepackage{makeidx}         
\usepackage{graphicx}        
\usepackage{multicol}        
\usepackage[bottom]{footmisc}

\makeindex             

\begin{document}

\title*{Synthesis models in a probabilistic framework: metrics of fitting}
\author{Miguel Cervi\~no\and
Valentina Luridiana}
\institute{Instituto de Astrof\'\i sica de Andaluc\'\i a (IAA-CSIC);
\texttt{mcs@iaa.es},
\texttt{vale@iaa.es}}
%
%
\maketitle

\begin{abstract}
Stellar population synthesis codes provide the description of the total luminosity resulting from the combination of an ensemble of sources as a function of given physical parameters. Their final aim is to enable users to make inferences about the physical parameters (star formation history, mass transformed into stars, etc.) from the light observed in stellar clusters or galaxies. However, synthesis codes results cannot be interpreted in an one-to-one relation between an observed total luminosity and a theoretical model result since (a) we have not access to the intimate composition of the ensemble of stars and (b) the very theoretical method implicit in populations synthesis makes use of (probabilistic) distributions and there is not an unique model result that describes the integrated luminosity of an ensemble. In general, synthesis models provide the {\it mean} value of the distribution of possible integrated luminosities, this distribution (and not only its mean value) being the actual description of the integrated luminosity.

Therefore, to obtain the closest model to an observation only provides confidence about the precision of such a fit, but not information about the accuracy of the result. In this contribution we show how to overcome this drawback and we propose the use of the theoretical mean-averaged dispersion that can be produced by synthesis models as a metric of fitting to infer accurate physical parameters of observed systems. 
\end{abstract}

\section{Introduction}
\label{sec:Intro}

Stellar populations synthesis models have been widely used to make inferences about the evolutionary status of observed systems by means of a comparison between observations and model results. In general it is assumed that the observed system is {\it well mixed} in terms of stellar populations and that models provide an one-to-one relation between the integrated luminosity and the physical conditions under the model is performed. In other words, it is assumed that the distribution of the possible integrated luminosities of a cluster with given physical conditions (including its number of stars) is a Dirac's delta-like distribution.

However, this assumption is not compatible with the very method implicit in synthesis codes, when they make use of probability distributions, like the Initial Mass Function, to describe the contribution of different stellar types to the integrated light. In the most optimistic case, and following the central limit theorem, the distribution of integrated luminosities must follow a gaussian like probability distribution in the asymptotic case of complete well mixed stellar types in the cluster (i.e. theoretical clusters with infinite number of stars). See \cite{CLCL06} for more details.

In this contribution we discuss the distributed nature of theoretical integrated luminosities of stellar clusters. We take advantage of such distributions to propose a metric that allows to establish the accuracy of fits of observations with models. We will assume that models are not affected by systematic errors (but see \cite{CLcancun,CLVO} for detailed discussions on this subject), and we will consider here only the intrinsic dispersion of the theoretical integrated luminosities.

\section{Understanding synthesis models}
\label{sec:SSPdesc} 

The first step needed to use population synthesis models is to understand how synthesis codes work. The main goal of synthesis codes is to obtain the integrated luminosity resulting from an ensemble of individual stars with given physical conditions (initial mass function IMF, metallicity $Z$, star formation history SFH, age $t$...). Let us assume that $n$ individual stars have been observed in a resolution element of a telescope (e.g. a pixel, a slit, an IFU, a set of several pixels, etc.). Since each star has an actual luminosity $\ell_{i}$, the total luminosity $L_{\mathrm{tot}}$ in the resolution element is:

\begin{equation}
L_{\mathrm{tot}} = \sum_{i=1}^n \ell_{i} = n\, \frac{1}{n} \sum_{i=1}^n \ell_{i} = n\, <\hat{\ell}>.
\label{eq:Ltot}
\end{equation}

\noindent where we have rewritten $L_{\mathrm{tot}}$ as a function of $n$ and $<\hat{\ell}>$ is the {\it estimation of the mean} luminosity of an individual star that represents the whole ensemble for the actual physical conditions.

However, we cannot perform such a sum and obtain the actual integrated luminosity of {\it this} set of individual stars since these stars are not resolved and we do not know the individual $\ell_{i}$ values. Instead, we can compute the {\it actual mean luminosity} of all the ensembles with the same physical conditions: It should be the composition of $N_{\mathrm{class}}$ different stellar populations (or different stellar types) for the given physical conditions. If we assume that a given stellar type ``$j$'' can be represented by a luminosity $l_{j}$ and we know {\it the probability} $w_{j}$ to have a star with a given stellar type determinated by the given physical conditions ($w_{j}=w_{j}[t, \mathrm{Z}, \mathrm{IMF}, \mathrm{SFH}]$), then a total luminosity of the system will be

\begin{eqnarray}
L_{\mathrm{tot}}^{\mathrm{theo}}(n) &=& n\, \sum_{j=1}^{N_{\mathrm{class}}} w_{j}\, l_{j} = n\,  <\ell>,\label{eq:LSSP}
\\
&&\sum_{j=1}^{N_{\mathrm{class}}} w_{j} = 1.\label{eq:LSSPnorm}
\end{eqnarray}

\noindent where $<\ell>$ is the luminosity of an {\it average} star that represents the ensemble for the given physical conditions. Synthesis codes provide $<\ell>$ with a suitable normalization, like the one in Eq. \ref{eq:LSSPnorm}.

Obviously, {\it the average} integrated luminosity of stellar clusters with $n$ stars is $n$ times the luminosity of this average star, but {\it the possible} integrated luminosity of a cluster is a distributed quantity. In fact, $L_{\mathrm{tot}}^{\mathrm{theo}}(n)$ in Eq. \ref{eq:LSSP} does not provide the whole description of the possible integrated luminosity of an observation of $n$ unresolved stars, but only the mean value of the underlying probability distribution that describes the possible integrated luminosities of all observations that contain $n$ unresolved stars. In the following we will use $<L_{\mathrm{tot}}^{\mathrm{theo}}(n)>$ instead of $L_{\mathrm{tot}}^{\mathrm{theo}}(n)$ to avoid confusion about this point.

Since we know that the theoretical integrated luminosity is a distributed quantity, we can go further and also obtain other quantities, besides the mean, which describe this distribution. In particular, the variance (which depends on how many unresolved stars are in the resolution element) and the variance normalized to the mean integrated luminosity (which is independent on the number of stars), we refer \cite{CLCL06} for more details in this subject\footnote{Eq. \ref{eq:SSPsbf} is related to Surface Brightness Fluctuations (SBF) when they are considered as a description of the variance of synthesis models; see \cite{Buzz89,Buzz93,Buzz06,CLJ07} for more details.}.

\begin{eqnarray}
\sigma^2[L_{\mathrm{tot}}^{\mathrm{theo}}(n)] &=& n \sigma^2(\ell) = n \, \left(\sum_{j=1}^{N_{\mathrm{class}}} w_{j}\, l_{j}^2 - <\ell>^2\right);\label{eq:SSPvar}\\
  & &
\frac{\sigma^2[L_{\mathrm{tot}}^{\mathrm{theo}}(n)]}{<L_{\mathrm{tot}}^{\mathrm{theo}}(n)>} = \frac{\sigma^2(\ell)}{<\ell>}.\label{eq:SSPsbf}
\end{eqnarray}

Summing up, {\it synthesis codes provide the luminosity of an average star or cluster} (depending on the normalization used in Eq. \ref{eq:LSSPnorm}) {\it that describes the mean luminosity of the ensemble for given physical conditions}. Codes can also provide the dispersion of the integrated luminosities around the mean by the use of Eq. \ref{eq:SSPvar} or Eq. \ref{eq:SSPsbf}.

In the following sections we show two different uses of the dispersion as  metrics for goodness-of-fit in stellar population studies. We will assume that the theoretical probability distributions of integrated luminosities follow a gaussian distribution (see \cite{CLCL06} for reliable criteria of gaussianity of synthesis model results).

\section{Studying stellar populations with synthesis models}

Let us move now to the application of synthesis models to study stellar populations in galaxies and stellar clusters. The problem is to infer the evolutionary conditions that are compatible with the observed integrated luminosity $L_{\mathrm{tot}}$ and to establish the confidence of the different hypothesis. Following D'Agostini work {\it From Observations to Hypotheses} \cite{dagostini}:  ``it becomes crucial to learn how to deal quantitatively with probability of causes because the {\it ``problem(s) in the probability of causes... may be said to be the essential problem(s) of the experimental method''} (Poincar\'e \cite{poincare}).''. 

\begin{figure}
\centering
\includegraphics[width=\textwidth]{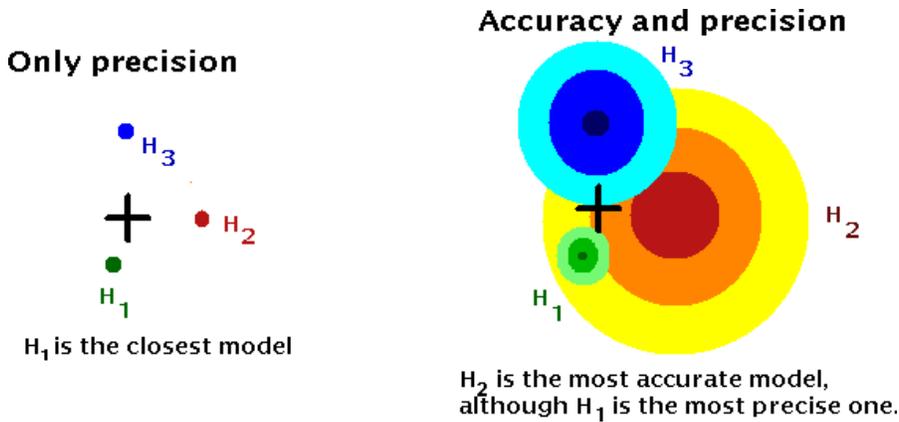}
\caption{Precision and accuracy; see text for details.}
\label{fig:PrevsAcu}       
\end{figure}

Let us illustrate the problem in Fig. \ref{fig:PrevsAcu}. We want to evaluate which model $\mathrm{H}_{1}$, $\mathrm{H}_{2}$ or $\mathrm{H}_{3}$ is more compatible with an observation (black cross). In the left panel we show the situation where only the mean value of the model is taken into account. In this case, $\mathrm{H}_{1}$ is the closest model to the observation and it is the ``best'' solution. In the right panel we show the situation when the complete theoretical distribution is taken into account. The different concentric circles represent 1, 2 and 3 $\sigma$ confidence intervals associated to {\it each one} of the models. $\mathrm{H}_{1}$ is still the closest model but the probability that the observation is compatible with $\mathrm{H}_{1}$ is lower than 1\% (i.e. the observation lies outside of the 99\% confidence interval of $\mathrm{H}_{1}$). The observation is compatible with $\mathrm{H}_{2}$ in a 2$\sigma$ confidence interval of $\mathrm{H}_{2}$ and compatible with $\mathrm{H}_{3}$ in a 3$\sigma$ confidence interval of $\mathrm{H}_{3}$. Hence, the ``closest'' model (following each model's metrics) is $\mathrm{H}_{2}$. This second solution is different than the previous (precise but not accurate) one.

Summing up, since our main interest is to have accurate (and not only precise) results, it is necessary to know the complete probability distributions of theoretical integrated luminosities and not only its mean values $<L_{\mathrm{tot}}^{\mathrm{theo}}(n)>$. The probability distribution of theoretical integrated luminosities provides a {\it metric for goodness-of-fit} to stellar population studies. Analysis method must obtain the closest model to the observation according with the metric defined by each model itself, and not by an ``absolute'' metric that does not take into account the intrinsic model dispersion.

We illustrate in the following sections two different ways of using the metric associated to theoretical models.
 
\section{Distance-dependent analysis}

Let us first work with the variance of the distribution ($\sigma^2$  in Eq. \ref{eq:SSPvar}). The dispersion ($\sigma$) of the integrated luminosity is proportional to the square-root of the number of stars in the observed population, so we need to have an estimation of how many stars are in our resolution element, or, equivalently, the distance to the source. 

\begin{figure}
\centering
\includegraphics[height=0.90\textwidth,angle=-90]{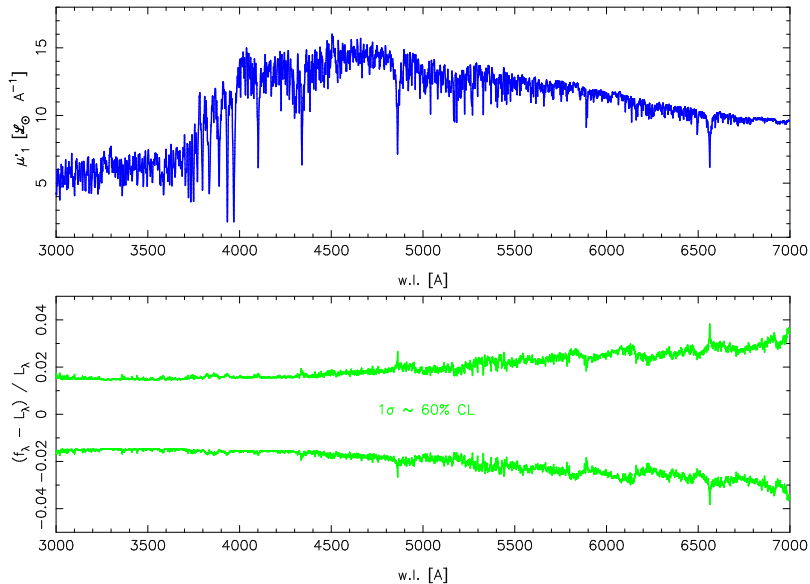}
\caption{{\it Top:} Integrated {\it mean} spectra of a solar metallicity 1 Ga old cluster with $10^5$ stars. {\it Bottom:} $1\sigma$ confidence interval of the relative dispersion of $(L_{\lambda} - <L_{\mathrm{tot}}^{\mathrm{theo}}(n=10^5)>) / <L_{\mathrm{tot}}^{\mathrm{theo}}(n=10^5)>$. Models from \cite{GDetal05}.}
\label{fig:SSPrelative}       
\end{figure}

\begin{figure}
\centering
\includegraphics[height=0.90\textwidth,angle=-90]{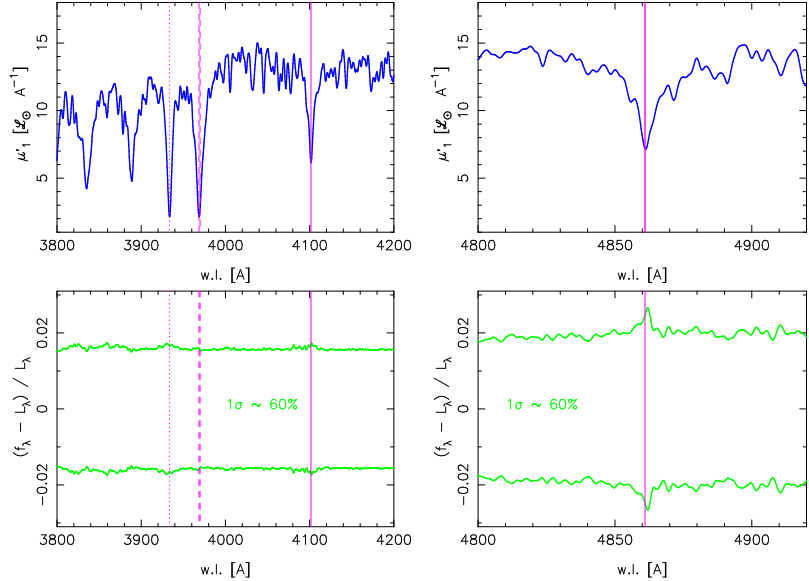}
\caption{Detail of Fig \ref{fig:SSPrelative} in the H$\delta$ wavelength range ({\it left}, with vertical lines showing the position of H$\delta$, Ca {\sc ii} H + H$\epsilon$ and Ca {\sc ii} K lines) and the H$\beta$ lines ({\it right}, with vertical line showing the position of H$\beta$ line).}
\label{fig:SSPrelative_detail}       
\end{figure}

Figs. \ref{fig:SSPrelative} and \ref{fig:SSPrelative_detail} show the mean energy distribution and relative dispersion of a model of a solar metallicity 1 Ga old cluster with $10^5$ stars. The first thing to notice is that the relative dispersion increases almost by a factor two (at this age and metallicity) at red wavelengths. This happens because the stars that contribute the most {\it to the mean} integrated luminosity at red wavelengths are also the most sparse ones and a small variation in the total number of such stars is translated to a large variation in the integrated luminosity. This situation can be reformulated by saying that infrared wavelengths have a lower {\it effective number} of stars than the visible (see \cite{Buzz89,Buzz93,Buzz06} for a more detailed description and implications about the effective number of stars that contibute to a given wavelength).

Second: line profiles have different dispersion degree (cfr. Fig. \ref{fig:SSPrelative_detail}). The H$\beta$ line has an {\it intrinsic} higher dispersion than H$\delta$.  This means that the profile of H$\beta$ is more sensitive than the profile of H$\delta$ to particular stellar populations in the cluster, so it can provide information about those particular stellar populations, but not about the {\it overall} population.

\section{Distance independent analysis}

\begin{figure}
\centering
\includegraphics[height=0.9\textwidth,angle=-90]{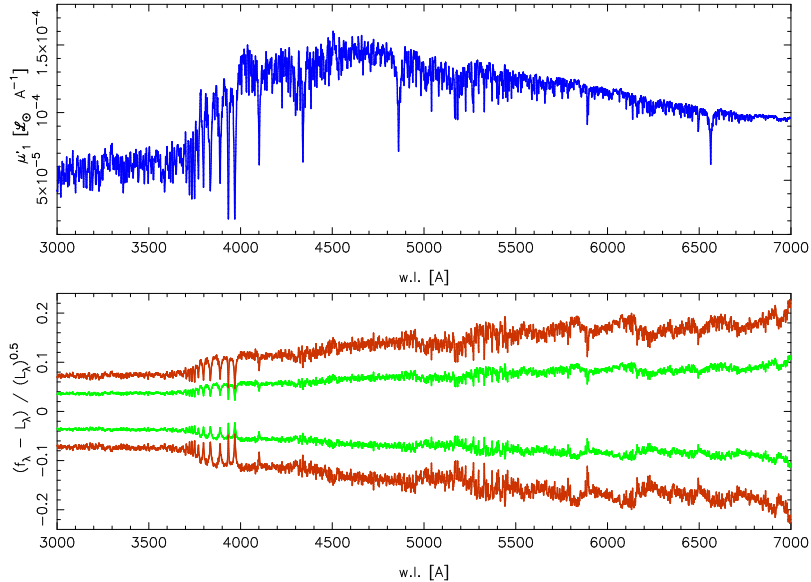}
\caption{{\it Top:} Integrated {\it mean} spectra of a 1 Ga old cluster. {\it Bottom:} 1 and 2$\sigma$ confidence intervals of the mean-averaged dispersion $(L_{\lambda} - <L_{\mathrm{tot}}^{\mathrm{theo}}>) / \sqrt{<L_{\mathrm{tot}}^{\mathrm{theo}}>}$. Models from \cite{GDetal05}.}
\label{fig:SSPsbf}       
\end{figure}

\begin{figure}
\centering
\includegraphics[height=0.9\textwidth,angle=-90]{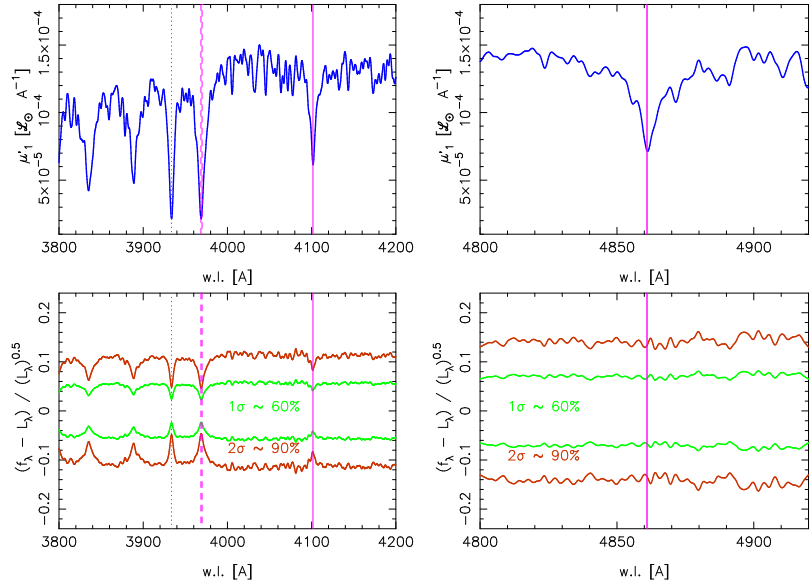}
\caption{Detail of Fig \ref{fig:SSPsbf} in the H$\delta$ wavelength range ({\it left}, with vertical lines showing the position of H$\delta$, Ca {\sc ii} H + H$\epsilon$ and Ca {\sc ii} K lines) and the H$\beta$ wavelength range ({\it right}, with vertical line showing the position of the H$\beta$ line).}
\label{fig:SSPsbf_detail}       
\end{figure}

As an alternative to the relative dispersion, we can use the mean-averaged dispersion defined in Eq. \ref{eq:SSPsbf}. The advantage of the mean-averaged dispersion is that it is independent of the number of stars in the cluster (hence, also independent of the distance). Figs. \ref{fig:SSPsbf} and \ref{fig:SSPsbf_detail} show the mean energy distribution and mean-averaged dispersion of a model of a solar metallicity 1 Ga old. In this case, the mean spectra (top panels) is the one of an average individual star, as mentioned in section \ref{sec:SSPdesc}.

As previously, the mean-averaged dispersion increases at red wavelengths. Absorption lines show a different behavior when expressed in this new units, but again, it shows similar conclusions: not all absorption lines can be fitted with the same precision.

\section{Conclusions}

Four conclusions can be drawn from this work: (a) the comparison of models and observations must be done taking into account the intrinsic metric of each model and not an absolute metric, (b) not all wavelengths must have the same weight in the fitting, (c) fits of line profiles have, in general, a larger dispersion than fits of the continuum, and (d) the more sensitive is the index (line profile, equivalent width, color...) about {\it a particular} stellar population, the less useful is the index to describe the {\it overall} evolutionary status of the population.

\section*{Acknowledgments}

This research has been supported by the Spanish MCyT and FEDER funds through the grant AYA2004-02703. MC is supported by a Ram\'on y Cajal fellowship.



\printindex
\end{document}